\documentclass[notitlepage,aps,prl,showpacs,superscriptaddress,twocolumn,amssymb,amsfonts]{revtex4-1}
\usepackage{times}
\usepackage{amsmath,bm}
\usepackage{graphics}
\usepackage{dsfont}
\usepackage{subfigure}
\usepackage{epsfig}
\usepackage{hyperref}
\usepackage{units}
\usepackage{color}

\begin{document}

\title{Efficient variational diagonalization of fully many-body localized Hamiltonians}

\author{Frank Pollmann}
\affiliation{\mbox{Max-Planck-Institut f\"ur Physik komplexer Systeme, N\"othnitzer Str.\ 38, 01187 Dresden, Germany}}
\author{Vedika Khemani}
\affiliation{\mbox{Department of Physics, Princeton University, Princeton, NJ 08544, USA}}
\affiliation{\mbox{Max-Planck-Institut f\"ur Physik komplexer Systeme, N\"othnitzer Str.\ 38, 01187 Dresden, Germany}}
\author{J. Ignacio Cirac}
\affiliation{\mbox{Max-Planck-Institut f\"ur Quantenoptik, Hans-Kopfermann-Str. 1, D-85748 Garching, Germany}}
\author{S. L. Sondhi}
\affiliation{\mbox{Physics Department, Princeton University, Princeton, NJ 08544, USA}}
\affiliation{\mbox{Max-Planck-Institut f\"ur Physik komplexer Systeme, N\"othnitzer Str.\ 38, 01187 Dresden, Germany}}

\begin{abstract}
We introduce a  unitary matrix-product operator (UMPO) based variational method that approximately finds \emph{all} the eigenstates of fully many-body localized (fMBL) one-dimensional Hamiltonians. The computational cost of the variational optimization scales linearly with system size for a fixed bond dimension of the UMPO ansatz. We demonstrate the usefulness of our approach by considering the Heisenberg chain in a strongly disordered magnetic field for which we compare the approximation to exact diagonalization results.
\end{abstract}

\maketitle

\noindent
{\bf Introduction:} The phenomenon of many-body localization (MBL) generalizes Anderson localization \cite{Anderson:1958vr} (AL) to interacting systems \cite{Fleishman:1980, Gornyi:2005,Basko:2006hh}.
In the Anderson problem the many-body Fock/Slater states constructed from the single
particle states have two important features. First, they exhibit an economical description---$L$ single particle states for
a system of size $L$ are sufficient to construct all $2^L$ many-body states. Second, all many-body states exhibit an 
area law for the entanglement entropy stemming from the localized nature of the constituent single particle states. Naturally,
attention has focused on what happens to these two properties in the MBL regime.

It was noted early on \cite{Pal:2010gr} that many-body eigenstates in the MBL regime would have only local entanglement and thus obey area laws  \cite{Srednick:1993we,Hastings:2007sd,Verstraete:2006em}. 
Subsequently Bauer and Nayak \cite{Bauer:2013jw} examined the behavior of the
entanglement entropy in detail and demonstrated the area law as well as deviations due to rare regions and states. 
In another set of papers \cite{Huse:2013uc, Huse:2014uy} the phenomenology of MBL systems was traced to an emergent set of $L$ commuting local integrals of motion (often called ``l-bits'') 
which are believed to exist in  fMBL systems---i.e. systems in which \emph{all} many-body eigenstates are localized.

These two developments invite a natural closure in which
the full set of $2^L$ many-body eigenstates are explicitly constructed from $O(L)$ local ingredients, at least approximately. 
The well known connection of the area law to matrix-product state (MPS)/tensor network representations of many-body states suggests that the latter are the correct language in which to carry
out this program. The program has two components: showing that such a compact representation exists and providing
a recipe for finding it without recourse to a knowledge of the exact eigenstates, potentially rendering a much larger
range of system sizes computationally tractable. 

In an important development, two groups have addressed the existence problem. Building on earlier work \cite{Pekker:2013vt}, Pekker and Clark (PC) \cite{Pekker:2014ux} have shown that the unitary operators that exactly diagonalize fMBL systems can be represented by matrix products operators (MPOs) \cite{Verstraete:2004}  of bond dimensions that appear to grow
very slowly with system size---in contrast to delocalized systems where the dimension grows exponentially with system size. The slow growth that they do observe is presumably due to rare many-body resonances/Griffiths effects; in its absence, the MPOs would 
yield the sought after $O(L)$ local description of the full spectrum. Parallel work  \cite{Chandran:2014} argued for the congruent result that the presence of local integrals of motion implies the existence of a single ``spectral tensor network'' that efficiently represents the entire spectrum of energy eigenstates in the fMBL phase. These developments however have not led to an
algorithm for finding a compact representation directly and even finding MPOs representing exactly known diagonalizing unitaries {\it a la} PC scales exponentially with system size \footnote{The  PC prescription matches eigenstates obtained via exact diagonalization (ED) to the  ``best'' (most local) diagonalizing unitary operator in a time that scales as $O(2^L)$ instead of the prohibitive worst case time  which scales as $O(2^L!)$. Although it is not yet clear how to exhaust all gauge degrees of freedom to find the optimal representation}.

\begin{figure}[tb]
  \includegraphics[width=\columnwidth]{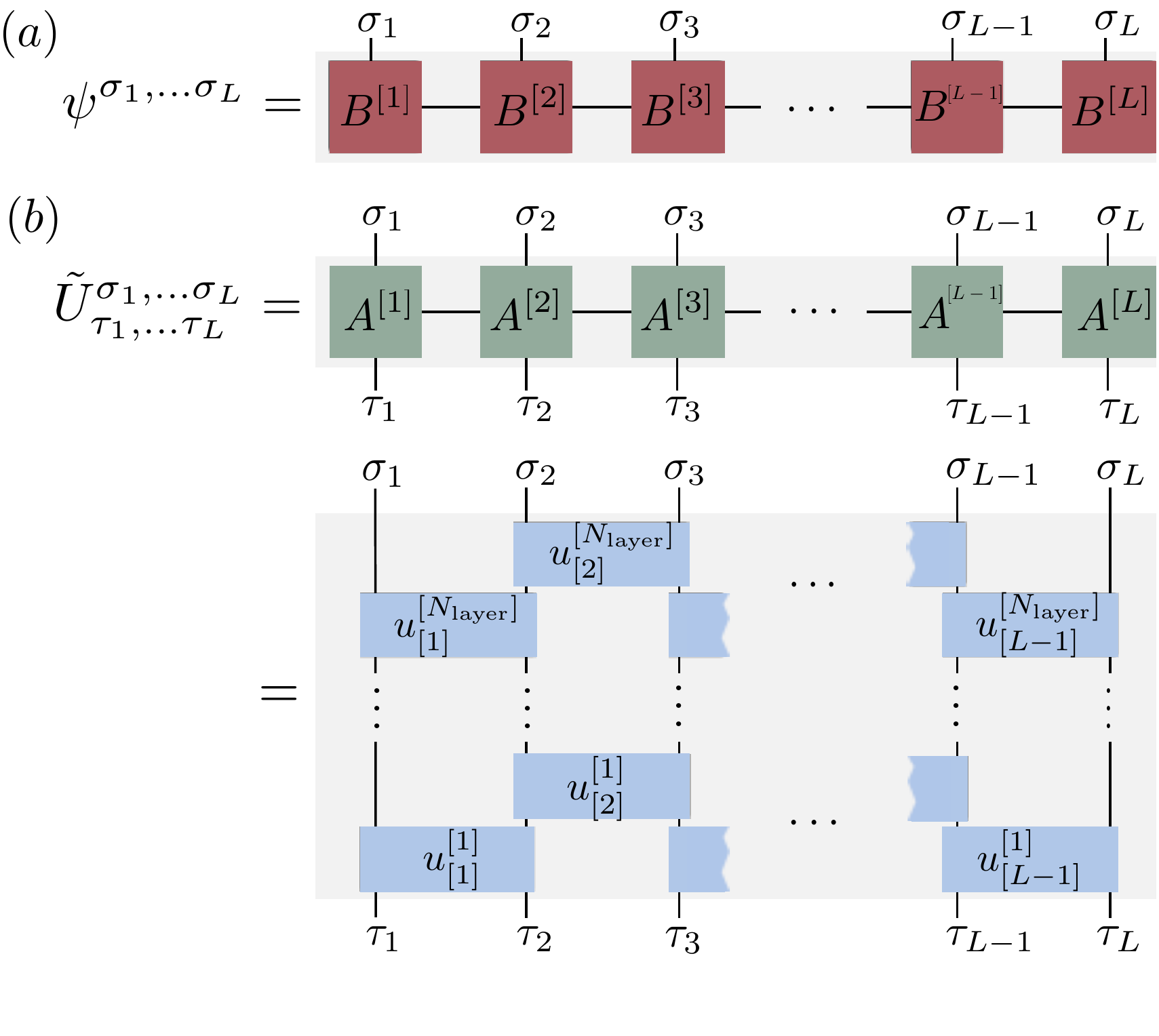} \\
  \caption{(a) Schematic representation of an MPS representation of a state $|\psi\rangle$. (b) Variational ansatz for the unitary $U$ that encodes \emph{all} eigenstates of a fully many-body localized Hamiltonian. The local unitaries $u^{[m]}_{[n]}$ are parametrized as  $u^{[m]}_{[n]} = e^{i S^{[m]}_{[n]}}$ with real symmetric matrices $S^{[m]}_{[n]}$, $n=1\dots L-1$ and $m=1\dots N_{\mathrm{layer}}$.\label{fig:vumpo}} 
 \end{figure}

In this paper we propose an approach to directly and efficiently find an approximate compact representation of the diagonalizing unitary by using a variational unitary MPO (VUMPO) ansatz. To this end, we construct a cost function whose minimum yields the exact unitary and, hence, the \emph{entire} set of $2^L$ exact eigenstates of a system of $L$ qubits.  We show that for a fixed bond dimension of the approximate $\tilde{U}$,  optimizing the cost-function in $d=1$ can be performed at a computational cost that is only \emph{linear} in system size which, in theory, allows us to access system sizes far beyond those possible by ED.

\noindent
{\bf MPS and MPO notation:} An MPS representation of a quantum state living in a basis spanned by $L$ qubits takes the form 
\begin{equation}
|\psi\rangle = \sum_{\{\boldsymbol{\sigma}\}} \sum_{0\le \gamma_i < D} B_{\gamma_1}^{[1] \sigma_1} B_{\gamma_1 \gamma_2}^{[2]\sigma_2} \cdots B_{\gamma_{L-1}}^{[L] \sigma_L} |\sigma_1  \cdots \sigma_L\rangle,
\end{equation}
 whereas an MPO representation of an operator in the same Hilbert space takes the form
\begin{equation}
O = \sum_{\substack {0\le \gamma_i < D \\ \{\boldsymbol{\sigma}\}, \{\boldsymbol{\tau}\}} } A_{\gamma_1}^{[1] \sigma_1, \tau_1} \cdots A_{\gamma_{L-1}}^{[L] \sigma_L, \tau_L} |\sigma_1  \cdots \sigma_L\rangle\langle\tau_1 \cdots \tau_L|,
\label{Eq:MPO}
\end{equation}
 where $\sigma_i, \tau_i \in {\{\uparrow, \downarrow\}}$ and we use a compact notation in which $\boldsymbol{{\sigma}}= \sigma_1, \sigma_2, \cdots, \sigma_L$ denotes the $2^L$ states (analogous for $\boldsymbol{{\tau}}$). Figure~\ref{fig:vumpo} shows a pictorial representation of these objects. The MPSs/MPOs are represented by rank three/four tensors $B^{[i]}/A^{[i]}$ on each site $i$ (except the first and last tensors which are rank two/three); the external leg(s) $\sigma_i, \tau_i$ refer to the physical spin indices whereas the $\gamma_i$ are the internal virtual indices that are contracted. Each $B^{[i] \sigma_i}/ A^{[i] \sigma_i \tau_i}$ is a $D^2$ dimensional matrix where $D$ is the bond-dimension of the matrix. 

\noindent
{\bf Method:}
We now introduce the VUMPO ansatz and an algorithm to numerially optimize it. Let us assume that $H$ is an fMBL Hamiltonian defined on an $L$-site chain of spin $1/2$ operators. 
It is our goal to find a unitary MPO approximation $\tilde{U}$ of the unitary that diagonalizes the Hamiltonian such that the $2^L$ eigenstates of $H$ are given by 
\begin{equation}
|\psi_{\boldsymbol{\tau}}\rangle \approx \sum_{\{\boldsymbol{\sigma}\}}\tilde{U}^{\boldsymbol{\sigma}}_{\boldsymbol{\tau}}|\boldsymbol{\sigma}\rangle. 
\end{equation}
In the parlance of Refs. \cite{Huse:2013uc, Huse:2014uy}, the physical basis operators $\sigma_i$ are the ``p-bits'' wheras the $\tau_i$ are the local ``l-bits''. Each eigenstate is labeled by the occupation of l-bits $\boldsymbol{\tau}  = \{\uparrow \uparrow \downarrow \cdots \uparrow\}$, and is obtained by acting with the MPO representation of $U$ on the product state $|\boldsymbol{\tau}\rangle$. 
In this language of MPOs, it is clear how the $2^L$ MB eigenstates are constructed from the $L$ matrices $A^{[i]\tau_i}$; further, if the bond-dimension of the matrices scales as $O(1)$ with the system size, the eigenstates are only locally entangled in the p-bit basis and a description of the full eigenbasis in terms of $O(L)$ \textit{local} ingredients is possible.

The VUMPO is found by numerically minimizing the cost functional
\begin{align}
f(\{A^{[n]}\}) =\sum_{\{\boldsymbol{\tau}\}}  \langle\psi_{\boldsymbol{\tau}}|H^2|\psi_{\boldsymbol{\tau}}\rangle - \langle\psi_{\boldsymbol{\tau}}|H|\psi_{\boldsymbol{\tau}}\rangle^2  \ge 0 \label{eq:cost},
\end{align}
with $\langle\psi_{\boldsymbol{\tau}}|\psi_{\boldsymbol{\tau'}}\rangle = \delta_{\boldsymbol{\tau},\boldsymbol{\tau'}}$. The cost function is the variance of the energy summed over all approximate MB eigenstates.  Naively, one might expect the time to evaluate the cost function Eq.~(\ref{eq:cost}) to scale exponentially with the system size $L$ as the sum is performed over $2^L$ MB eigenstates. However, remarkably, the computation can be performed in a time scaling \textit{linearly} with system size \cite{Verstraete:2004}! For example, the term $\sum_{\{\boldsymbol{\tau}\}} \langle\psi_{\boldsymbol{\tau}}|H|\psi_{\boldsymbol{\tau}}\rangle^2$ can be evaluated by ``doubling'' the degrees of freedom and defining a state $|\phi\rangle = \sum_{\{\boldsymbol{\tau}\}}  |\psi_{\boldsymbol{\tau}}\rangle|\psi_{\boldsymbol{\tau}}\rangle|\boldsymbol{\tau}\rangle$. With this notation we find that $\sum_{\{\boldsymbol{\tau}\}}  \langle\psi_{\boldsymbol{\tau}}|H|\psi_{\boldsymbol{\tau}}\rangle^2 = \langle\phi|H\otimes H \otimes \mathds{1}|\phi\rangle$. This expectation value can be efficiently evaluated using the MPO formalism and the most expensive part of the evaluation scales, for a given Hamiltonian in MPO form, as $\propto LD^5$ (see Appendix A for details and a diagrammatic representation of the terms). 
One can now iteratively minimize $f$ by locally optimizing each $A^{[n]}$  using, for example, the conjugate gradient algorithm.

In general, an MPO compression of a unitary operator will not strictly respect unitarity. To get a valid positive-definite cost function in these cases, we need to add a Lagrange multiplier to enforce unitarity (or consider other cost functions which don't assume orthonormality of the eigenstates). In practice, these methods lead to either very unstable, or very computationally expensive optimizations. 

\begin{figure}[tb]
  \includegraphics[width=\columnwidth]{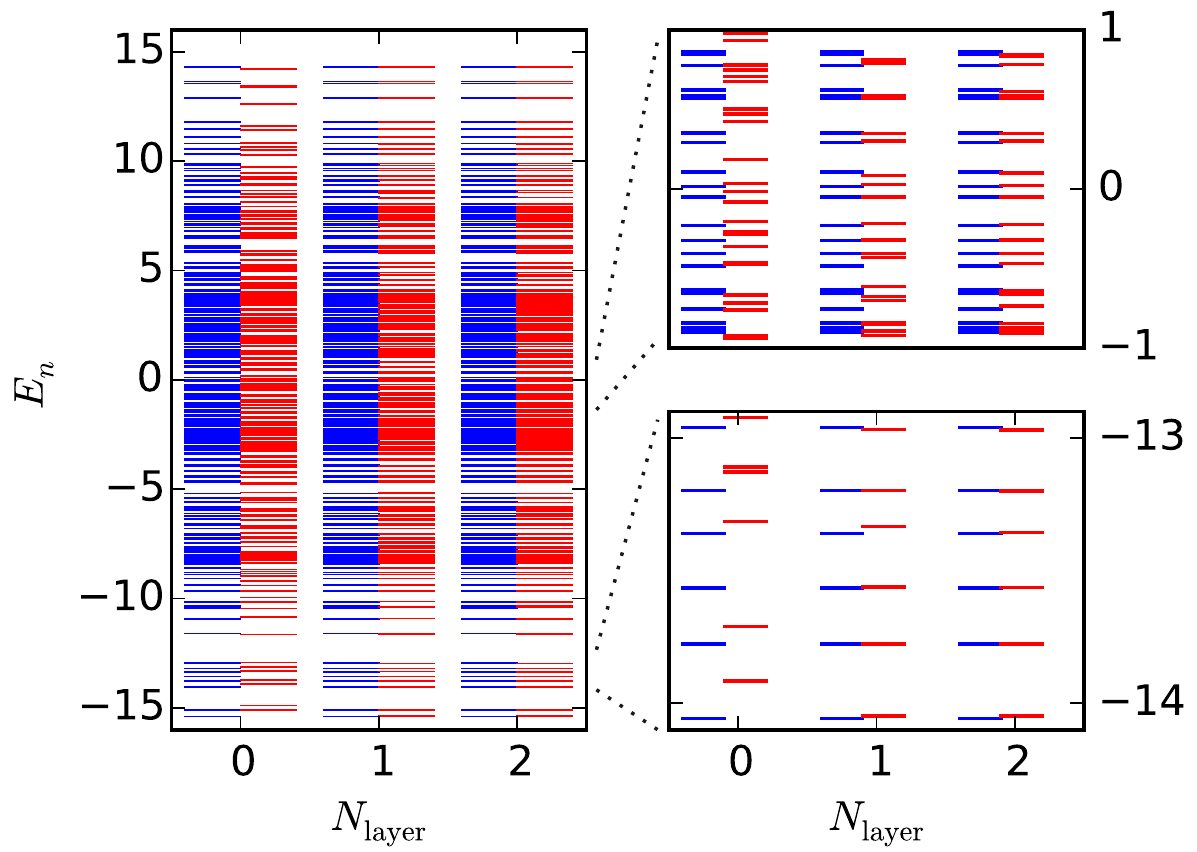} \\
  \caption{Comparison of the exact energy levels (blue lines)  with the ones found by the variational optimization (red lines) for $W=8$ and $L=8$ as a function of the number of layers of two-site gates. The right panel shows a zoom of some energy levels at the bottom and in the center of the spectrum.\label{fig:spectrum}}
 \end{figure}

The key to a stable optimization protocol turns on restricting our algorithm to the manifold of strictly unitary MPOs of a given bond-dimension. 
To achieve this, we parameterize the VUMPO as a finite depth circuit of two-site unitaries as shown in Fig.~\ref{fig:vumpo}(b). 
This Ansatz incorporates two important properties: (i) The VUMPO is unitary for all parameters and (ii) it is local for any finite $N_{\mathrm{layer}}$. We use a single unitary to obtain all eigenstates, but readers will note the obvious connection to the quantum computational notion \cite{Bauer:2014} 
that each MBL eigenstate can be constructed from a reference Fock state via the operation of a, in general different, finite 
depth circuit made up of local unitaries. 
Finally, we note that we can rewrite the unitary network as a strictly unitary MPO with bond dimension $D \le 2^{2N_{\mathrm{layer}}}$, where $N_{\rm layer}$ is the number of layers of two-site gates\footnote{The maximum bond dimension follows from the fact that the maximum entanglement for a bipartition of an approximate eigenstate is bounded by the number of two layer gates extending across the entanglement cut.}. However, this step is not necessary and we can evaluate the cost function by directly contracting the  unitaries circuit which, in fact, gives a considerable speed up for the systems we consider here \footnote{Using the locality of the unitary circuit, the cost function can be evaluated locally and thus it is, in principle, possible to generalize the approach to higher dimensions.}. 
 
The algorithm to find the VUMPO is then similar in spirit to the density matrix renormalization group (DMRG) method \cite{White:1992}, except instead of finding the lowest energy state, we minimize the cost function Eq.~(\ref{eq:cost}) by sweeping through the local unitaries:\\
\noindent(i) Initialize the local unitaries $u^{[m]}_{[n]} = e^{i S^{[m]}_{[n]}}$ by choosing random symmetric matrices $S_{[n]}^{[m]}$, where $n = 1, 2, \cdots L$ and $m = 1,2, \cdots N_{\rm layer}$. \\
\noindent(ii)  Locally minimize the cost function by varying the elements of a given $S_{[n]}^{[m]}$ by using, e.g., a conjugate gradient method. \\
\noindent(iii) Update the network and repeat the previous step for the next unitary.\\
\noindent(iv) Continue the sweeping procedure by minimizing the local unitaries successively until convergence. A full sweep across all the unitaries has to scale as $O(L)$. \\
We find that the number of steps needed for convergence appears to be approximately independent of $L$. This gives an overall scaling of the algorithm as $O(LD^5) \sim O(L e^{N_{\rm layer}})$. Once the algorithm has converged, the VUMPO can be used to obtain all the eigenstates of the system, and to efficiently compute observables using the MPS formalism.

\noindent
{\bf Results:} We consider the Heisenberg model with random $z$-directed magnetic fields:
\begin{equation}
H = J \sum_n  \vec{S}_n\cdot\vec{S}_{n+1} - \sum_n h_n S_n^z. \label{eq:ham}
\end{equation}
where $\vec{S_n}$ are spin 1/2 operators and the fields $h_n$ are drawn randomly from the interval $[-W,W]$ and we set $J=1$. 
This model has been studied extensively in the context of MBL and several numerical studies strongly suggest that $H$ is  fMBL for $W \gtrsim 3.5$ \cite{Pal:2010gr,Bardarson:2012gc,Luitz:2015}.  
  
\emph{Energy Spectrum: }We begin by comparing the energies obtained using the VUMPO approach with the exact spectrum (full diagonalization).
The converged results for $W = 8$ and $L=8$ with different numbers of layers $N_{\mathrm{layer}}$ are shown in Fig.~\ref{fig:spectrum}.
For $N_{\mathrm{layer}}=0$, the VUMPO is the identity (i.e, no variational parameters) and the resulting approximate eigenstates are simple product states of the form $|\sigma_1\rangle|\sigma_2\rangle\dots|\sigma_L\rangle$ with $\sigma_n=\uparrow,\downarrow$. The overall bandwidth in this case agrees relatively well with the exact results because $W$ is the dominant energy scale in the problem. 
However, as shown in the zoomed in plots, the deviation of individual energy levels is relatively large compared to the mean-level spacing because the product states completely neglect local quantum fluctuations which are present in the exact eigenstates. Increasing $N_{\mathrm{layer}}$ strongly improves the agreement between the exact and approximate energy levels as the network successively adds entanglement over longer distances.

 \begin{figure}[tb]
  \includegraphics[width=1.\columnwidth]{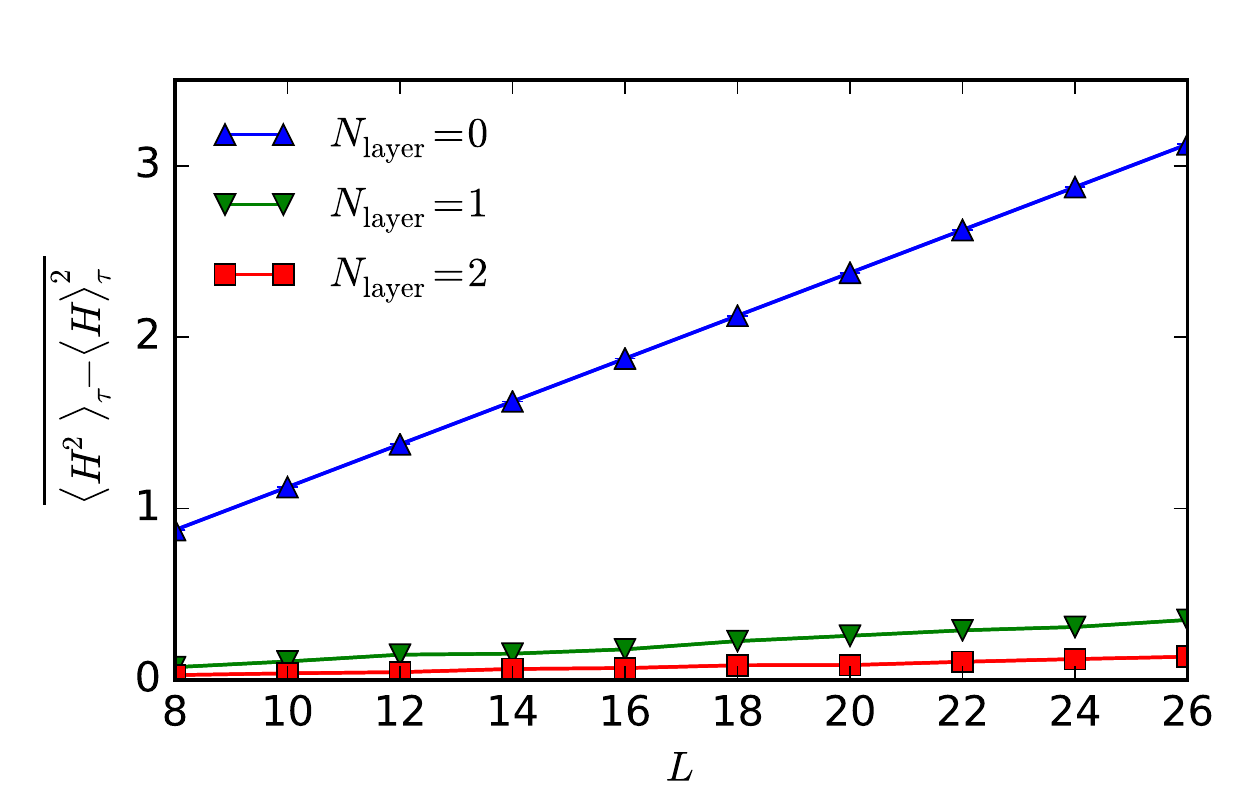} \\
  \caption{Mean variance of the energy as a function of system size for different number of layers for $W=8$.\label{fig:cost}}
 \end{figure}
 
Next we turn to the mean variance of the energy, which is simply the disorder averaged cost function Eq.~(\ref{eq:cost}) divided by $2^L$.  Figure~\ref{fig:cost} shows this quantity disorder averaged over 50 realizations as a function of system size for different fixed $N_{\mathrm{layer}}$. 
We observe a linear increase of the mean variance with system size, 
and find that the slope decreases as $N_{\mathrm{layer}}$ is increased. This tells us that for a given $N_{\mathrm{layer}}$ 
our approximate eigenstates involve a constant error per unit length which decreases as $N_{\mathrm{layer}}$ is increased.
This scaling is entirely exactly the same as that obtained for ground states obtained via DMRG.

\begin{figure}[tb]
\includegraphics[width=1.\columnwidth]{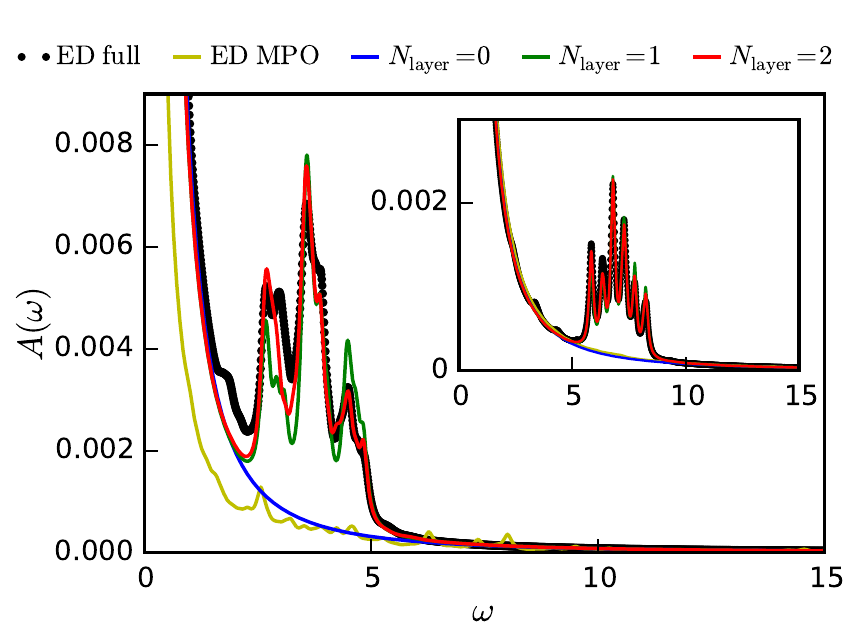}\\
\caption{Comparison of the exact spectral function $A(\omega)$ (black dots) with those obtained using different approximations (see text for details) for $L = 10$ and $W = 8$.  Spectra are shown using a Lorentzian broadening with an imaginary part of $\epsilon = 0.1$. Inset: Same data with $W = 16$. 
\label{fig:A}}
 \end{figure}
 
\emph{Spectral Functions:} To examine the quality of our approximated eigenstates (with a view to capturing local properties),  we use the VUMPO ansatz to obtain the infinite-temperature spectral function 
\begin{equation}
A(\omega) = \frac{1}{2^L}\sum_{\{\boldsymbol{\tau_1}\},\{\boldsymbol{\tau_2}\}}|\langle\boldsymbol{\tau_1}|S^z_{L/2}|\boldsymbol{\tau_2}\rangle|^2\delta(\omega - E_{\boldsymbol{\tau_1}} +  E_{\boldsymbol{\tau_2}}).
\end{equation}
Spectral functions can again be efficiently evaluated using matrix-product techniques and it is also possible to efficiently target different energy densities by considering finite-temperature spectral functions \cite{Verstraete:2004,Zwolak:2004qa}. 
Figure~\ref{fig:A} compares the $A(\omega)$ obtained using the VUMPO approach for $L=10$ with different disorder strengths and $N_{\mathrm{layer}}=0,1,2$ with the exact results. 
The spectral functions are dominated by a large peak at $\omega=0$ which reflects the strongly localized nature of the eigenstates, i.e., the eigenstates of $H$ are close to being eigenstates of local $S^z$ operators. It is interesting to compare the peaks at $\omega>0$ which are due to local fluctuations in the eigenstates.  Clearly,  $N_{\mathrm{layer}}=0$ does not show any features because the VUMPO is diagonal in $S^z$. When additional layers of unitaries are taken into account, the peak structure of $A(\omega)$ is well approximated. The agreement in both the frequencies and the intensities rapidly improve with increasing $N_{\rm layer}$, and the results match almost perfectly for $W = 16$. Note that despite the extremely strong disorder, simply approximating the eigenstates as product states fails to capture any of the interesting features. 

\emph{Comments on accuracy:} We have presented some evidence above for the accuracy of the VUMPO obtained by our
method. It remains to establish more precise theorems on what values of $N_{\mathrm{layer}}$ it would take to calculate various physical
quantities to a specified accuracy. In a step in that direction, PC have looked at the bond dimension needed to ensure that the smallest singular value in the Schmidt decomposition across any cut in $U$ is less than some fixed $\epsilon$. This ensures that the discarded weight on truncating the unitary to bond dimension $D$ is small. They found a slow growth of the $D_{\rm min}$ needed to achieve a desired $\epsilon$ with $L$. In the absence of rare resonances or Griffiths regions,  $D_{\rm min}$ would
presumably saturate at a fixed $O(1)$ value for a fixed error density independent of system size implying that we would need only $O(1)$ layers to represent the entire spectrum to the desired accuracy. As it is, with the resonances/Griffiths regions present, $D_{\rm min}$ is expected to grow as $\mathrm{poly}(L)$ whence $N_{\mathrm{layer}}$ will grow logarithmically. We should note however, that the PC criterion is 
not without its problems for the truncation
they would employ causes loss of unitarity. For example, let us return to our spectral function computation above but this time
we first obtain the  exact $2^L\times2^L$ dimensional unitary that diagonalizes $H$ and then compress it  to an MPO of a 
given bond dimension $D$. We do this by iteratively maximizing the ``overlap'' of an MPO with a fixed bond dimension  with the ``most local'' diagonalizing unitary obtained by following the PC prescription [16].
Because the compression scheme only minimizes the distance between the exact and the approximated unitary with respect to some operator norm, unitarity is not necessarily preserved.
As seen in Fig.~\ref{fig:A} (labeled ED MPO), when compressing $U_{\rm PC}$ to $D=16$ (which can exactly represent our $N_{\mathrm{layer}}=2$ results), the spectral functions $A(\omega)$ are very poorly reproduced. A reasonable agreement is only achieved for very large bond dimensions when the truncation error becomes negligible. 

\textbf{Summary and discussion:} 
We have introduced an algorithm to find a variational unitary MPO that approximately diagonalizes fully many-body localized Hamiltonians. Our method finds an approximation to all $2^L$ eigenstates of the Hamiltonian in a time that remarkably scales only linearly with system size! 
We have benchmarked the method by comparing the results to exact diagonalization for small systems and studied the scaling of the mean variance as a function of system size.
For a Heisenberg model in a strongly disordered field we find good qualitative and quantitative agreement of the obtained energies and spectral functions for a fixed $N_{\rm layer}$ and, importantly, rapid improvement with increasing $N_{\rm layer}$. 
With this work we have provided a proof of principle that we can efficiently (i.e, polynomially in system size) perform a variational calculation that provides a complete diagonalization of fMBL systems.
As the VUMPO encodes the entire set of eigenstates for fMBL Hamiltonians, many relevant observables such as spectral functions and conductivities can be evaluated efficiently at zero and finite  temperatures. 

A few comments are in order. First, it is intuitively clear that our VUMPOs should capture most of the structure of the eigenfunctions, or equivalently l-bits, out to a fixed ``lightcone'' radius, set by $N_{\rm layer}$. 
In terms of the dynamics this should allow accurate inclusion of local excitations on the same length scale and via the recently discussed connection between the energy and size of
many-body resonances \cite{sarangetal} down to a related frequency scale. Indeed, this feature can be effectively used to study different ``slices'' of the response function as more layers are added. For example, Figure~\ref{fig:A}  shows that the exact solution in the case of $W=8$ shows certain features at lower frequencies which are absent in the variational solution. Second, for a given VUMPO, one can construct\cite{Swingle:2013vl} a family of parent Hamiltonians $H = U^{\dagger}H^{\mathrm{diag}}U$ with the same eigenstates by picking different energy distributions for diagonal Hamiltonians in the  ``l-bit'' basis,  $H^{\mathrm{diag}}$. 

Going forward we can visualize many possible avenues for improving our method. 
Initially it may be possible to choose the same number of two-qubit gates in a different
architecture  \cite{ciracgates,Lamata:2008ab} to get a softer cutoff on the entanglement.
More ambitiously
we could allow for some two-qubit gates with a longer range and optimize over {\it both} the architecture of the unitary network, and the particular gates used. It is also possible to engineer the cost function to target a desired energy density via a pseudo-thermal weighting which could improve such focused results for fixed resource use and also allow MBL systems exhibiting mobilty edges to be treated. Of course the most desired improvement would be to run at $N_{\rm layer} > 2$ which is currently stymied 
by the exponential scaling of the cost function. As the diagrams to be contracted now start resembling 2D tensor-network graphs,  algorithms from this field could presumably be used to improve the scaling of contraction times. 

We thank Bryan Clark for useful comments on the manuscript. This work was supported by NSF Grant No. 1311781 and the John Templeton Foundation (VK and SLS) and the Alexander von Humboldt Foundation and the German Science Foundation (DFG) via the Gottfried Wilhelm Leibniz Prize Programme at MPI-PKS (SLS). 

%

\clearpage
\appendix
\section{Efficient evaluation of the cost functional}

In this section we discuss some details of how to efficiently evaluate the cost function Eq.~(\ref{eq:cost}) using the MPO formalism.
Due to the unitarity of $U$, the first term,  $\sum_{\boldsymbol{\tau}} \langle\psi_{\boldsymbol{\tau}}|H^2|\psi_{\boldsymbol{\tau}}\rangle$, is simply $\mathrm{Tr } H^2$.
If $H$ is represented by a $\chi$ dimensional MPO, the trace can by evaluated with a cost scaling as $\sim Ld^3\chi^2$ as shown in Fig.~\ref{fig:cost} (top); $d$ is the dimension of the local Hilbert space on each site and is equal to 2 for the spin-1/2 operators considered in this work. 

  \begin{figure}[tb]
  \includegraphics[width=1.\columnwidth]{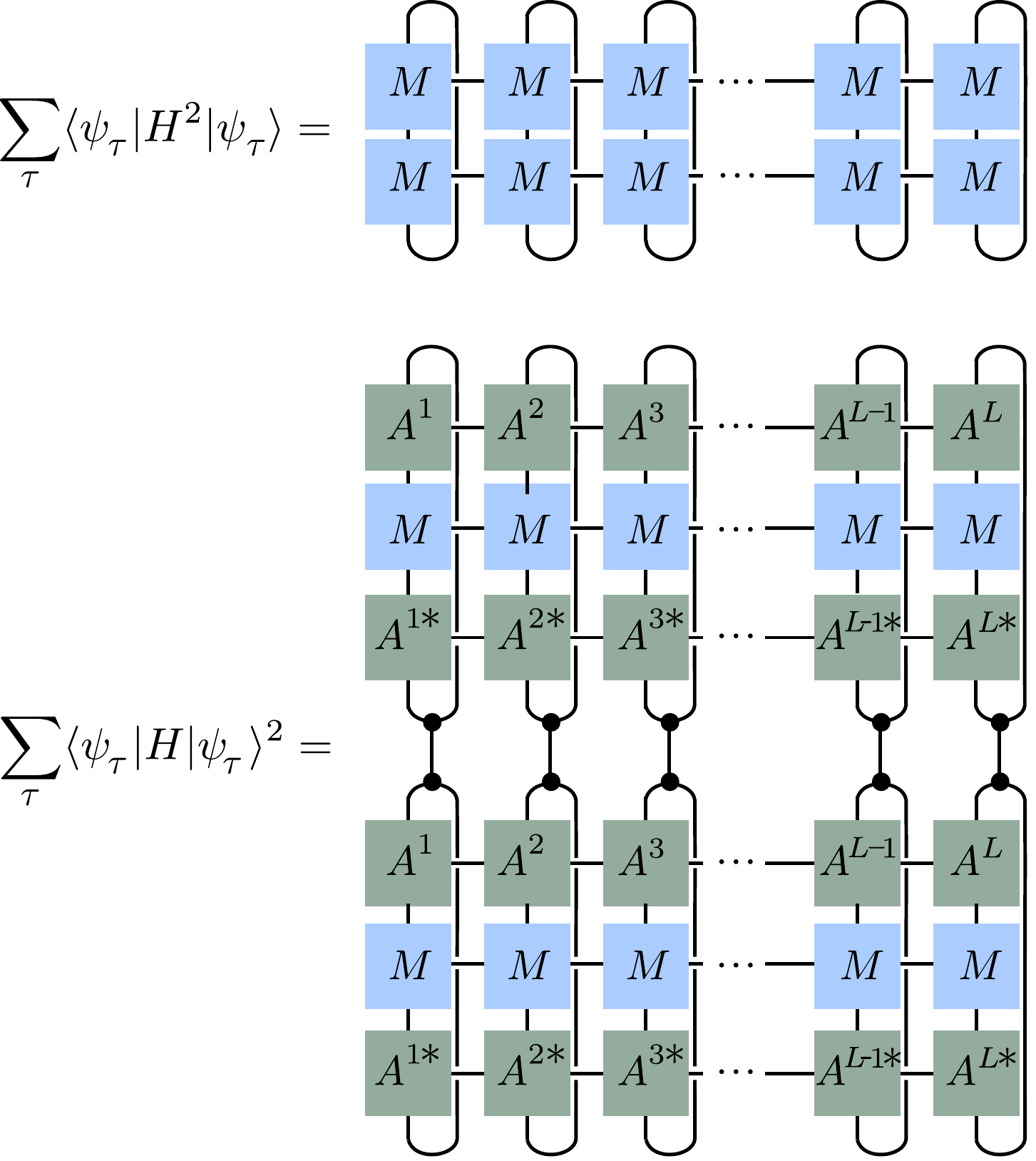} \\
  \caption{Diagrammatic representation of the tensor contractions required to evaluate the terms in the cost function Eq.~(\ref{eq:cost}). The tensors $A^{[n]}$ represent the unitary and $M$ the Hamiltonian. The back dots are delta functions $\delta_{a,b,c}$. \label{fig:cost} }
 \end{figure}

The second term, $\sum_{\boldsymbol{\tau}} \langle\psi_{\boldsymbol{\tau}}|H|\psi_{\boldsymbol{\tau}}\rangle^2$, is somewhat more challenging.
We first ``double'' the system by taking two identical copies and form a tensorproduct with a state $|\boldsymbol{\tau}\rangle$ (which is simply a product state of the ``l-bits''),
\begin{equation}
|\phi\rangle = \sum_{\boldsymbol{\tau}} |\psi_{\boldsymbol{\tau}}\rangle|\psi_{\boldsymbol{\tau}}\rangle|\boldsymbol{\tau}\rangle.
\end{equation} 
Using the state $|\phi\rangle$ and that $\langle\boldsymbol{\tau}|\boldsymbol{\tau'}\rangle=\delta_{\boldsymbol{\tau},\boldsymbol{\tau'}}$, we find that
\begin{equation}
\sum_{\boldsymbol{\tau}} \langle\psi_{\boldsymbol{\tau}}|H|\psi_{\boldsymbol{\tau}}\rangle^2 = \langle\phi|H\otimes H \otimes \mathds{1}|\phi\rangle.
\end{equation}
This expectation value can again be evaluated efficiently using the MPO formalism as demonstrated in Fig.~\ref{fig:cost} (bottom).
 Given that $D> \chi > d$, the most expensive part of the contraction scales as $\sim LD^5\chi^2d^4$.

\end{document}